\documentclass[12pt,preprint]{aastex}
\usepackage{epsfig}
\usepackage{graphicx}
\usepackage{epstopdf}
\begin{document}
\title{Deep Spitzer observations of infrared-faint radio sources: high-redshift radio-loud AGN?}
%\centerline{DRAFT: \today}
%\slugcomment{Will be submitted to ApJ, draft \today}
\floatsep=5mm
\intextsep=5mm
\dblfloatsep=5mm
\textfloatsep=5mm
\belowcaptionskip=5mm

 \author{
Ray P.~Norris,$^{\! 1}$
Jose~Afonso,$^{\! 2,3}$
Antonio~Cava,$^{\! 4}$
Duncan~Farrah,$^{\! 5}$
Minh~T.~Huynh,$^{\! 6}$
R.\,J.~Ivison,$^{\! 7,8}$
Matt~Jarvis,$^{\! 9}$
Mark~Lacy,$^{\! 10}$
Minnie~Mao,$^{\! 1,6,11,12}$
Claudia~Maraston,$^{\! 13}$
Jean-Christophe~Mauduit,$^{\! 6}$
Enno~Middelberg,$^{\! 14}$
Seb~Oliver,$^{\! 5}$
Nick~Seymour$^{15}$ and
Jason~Surace$^{6}$
}
\affil{
$^1$CSIRO Australia Telescope National Facility, PO Box 76, Epping, NSW, 1710, Australia,
email: {\tt Ray.Norris@csiro.au}\\
$^2$ Observat\'{o}rio Astron\'{o}mico de Lisboa, Faculdade de Ci\^{e}ncias, Universidade de Lisboa, Tapada da Ajuda,
1349-018 Lisbon, Portugal\\
$^3$ Centro de Astronomia e Astrof\'{\i}sica da Universidade de Lisboa, 1349-018 Lisbon, Portugal\\
$^4$ Departamento de Astrof\'{\i}sica, Facultad de CC. F\'{\i}sicas, Universidad Complutense de Madrid, E-28040 Madrid, Spain\\
%$^5$ Departamento de Astrofisica, Universidad de La Laguna, E-38205, La Laguna, Tenerife, Spain\\
$^5$Astronomy Centre, Dept. of Physics \& Astronomy, University of Sussex, Brighton BN1 9QH, UK\\
$^6$Infrared Processing and Analysis Center, MS220-6, California Institute of Technology,
Pasadena CA 91125, USA\\
$^7$UK Astronomy Technology Centre, Royal Observatory, Blackford Hill, Edinburgh EH9 3HJ, UK\\
$^8$Institute for Astronomy, University of Edinburgh, Blackford Hill, Edinburgh EH9 3HJ, UK\\
$^{9}$Centre for Astrophysics, Science \& Technology Research Institute, University of Hertfordshire, Hatfield, Herts, AL10 9AB, UK\\
$^{10}$NRAO, 520 Edgemont Road, Charlottesville, VA 22903, USA\\
$^{11}$School of Mathematics and Physics, University of Tasmania, Private Bag 37, Hobart, 7001, Australia\\
$^{12}$Anglo-Australian Observatory, PO Box 296, Epping, NSW, 1710, Australia\\
$^{13}$Institute of Cosmology and Gravitation, Dennis Sciama Building, Burnaby Road, Portsmouth, PO1 3FX, UK\\
$^{14}$Astronomisches Institut, Ruhr-Universit\"at Bochum, Universit\"atsstr. 150, 44801 Bochum, Germany\\
$^{15}$Mullard Space Science Laboratory, UCL, Holmbury St Mary, Dorking, Surrey, RH5 6NT, UK\\
}

\begin{abstract}

Infrared-faint radio sources (IFRSs) are a rare class of object which are relatively bright at radio wavelengths but very faint at  infrared and optical wavelengths. Here we present sensitive near-infrared observations of a sample of these sources taken as part of the  {\it Spitzer} Extragalactic Representative Volume Survey (SERVS). Nearly all the IFRSs are undetected at a level of $\sim 1 \mu$Jy in these new deep observations, and even the detections are consistent with confusion with unrelated galaxies.  A stacked image implies that the median flux density is  $S_{3.6\mu m} \sim 0.2$ \,$\mu$Jy or less, giving extreme values of the radio-infrared flux density ratio. Comparison of these objects with known classes of object suggests that the majority are probably  high-redshift radio-loud galaxies, possibly suffering from significant dust extinction.

\end{abstract}


\section{Introduction}

Infrared-faint radio sources (IFRSs) are a rare class of object which are relatively strong at radio wavelengths but very faint at infrared (IR) and optical wavelengths. They were first categorised in the Australia Telescope Large Area Survey
\citep[ATLAS:][]{norris2006} as radio sources
%brighter than a few hundred\,$\mu$Jy at 20 cm,
with no
observable IR counterpart in the co-spatial {\it Spitzer} Wide-area IR Extragalactic Survey \citep{lonsdale2004}.  Most have flux densities of a few hundred\,$\mu$Jy at 20 cm, but some are as bright as 20\,mJy.
%More specifically, IFRSs may be defined as objects with
%S$_{20\,cm}>100\,\mu$Jy and S$_{3.6\,\mu{\rm m}}<5\,\mu$Jy, although most IFRS have much higher 20\,cm flux densities, giving them flux density ratios
%S$_{20\,cm}/S_{3.6\,\mu{\rm m}} \gtrsim 400$.
They may be related to the optically invisible radio sources found by \cite{higdon2005, higdon2008}, which are compact radio sources with no optical counterpart to $R \sim 25.7$, although the IFRSs seem even more extreme than the Higdon objects.
\cite{norris2006} and \cite{middelberg2008a} have identified 51 such sources out of 2002 radio sources in the ATLAS survey. 

So far, four samples of IFRS have been identified: (a) the original ATLAS/SWIRE samples identified by \cite{norris2006} and \cite{middelberg2008a}, from which the sources in this paper are drawn, (b) the sample in the Spitzer First-Look Survey, identified by \cite{garn2008}, (c) the sample in ELAIS-N1 identified by \cite{grant2011}, (d) a sample in the COSMOS field \citep{Scoville2007} identified by \cite{Zinn2011}. These samples imply a sky density of of  $\sim$7 per deg$^2$ for $S_{\rm 20\,cm} > 0.1$\,mJy.

At the time of their discovery, the IFRS sources were unexpected, as SWIRE was thought to be deep enough to detect all extragalactic radio sources at z$\lesssim$2, regardless of whether star formation or active galactic nuclei (AGN) powered the radio emission.
Possible explanations were that these sources are i) high-redshift radio-loud AGN, ii) very obscured radio galaxies at more moderate redshifts ($1 < z < 2$), iii)
lobes of nearby but unidentified radio galaxies, iv) very obscured, luminous starburst galaxies, such as high-redshift submillimetre-selected galaxies \citep[SMGs --][]{smail97}, v) high-latitude pulsars, or vi) an unknown type of object. Of course it is also possible that they do not constitute a homogeneous class and harbour examples of some or all of the above.

The nature of IFRSs has been hard to determine because nearly all the information on them has been obtained at radio wavelengths. Spectroscopy is difficult because the hosts are optically faint and the radio positions can also have uncertainties of the order of a few arcsec. \cite{norris2006} stacked the positions of 22 IFRSs in the {\sl Spitzer} 3.6\,$\mu$m IRAC images and found no detection in the averaged image, showing that they are well below the SWIRE detection threshold. This was a surprising result at the time, as it was expected that the IFRSs represented the tail of a distribution reaching just below the SWIRE detection threshold, but it was confirmed by  \citet{garn2008} and by the new data presented here.

\cite{middelberg2008b} and \cite{norris2007} targeted six IFRSs with the Australian Long Baseline Array (LBA) and detected two of the sources.  
The \cite{norris2007} LBA detection constrained the source size to less than 0.03\,arcsec, suggesting a compact radio core, powered by an AGN.
\cite{middelberg2008b} found the size and radio luminosity of their LBA-detected source to be consistent with a high-redshift ($z > 1$) compact, steep-spectrum (CSS) source.
The VLBI detections rule out the possibility that these particular IFRSs are simply the radio lobes of unidentified radio galaxies, or star-forming galaxies, though the initial VLBI targets were inevitably amongst the most radio-bright examples in the sample.

\cite{garn2008} stacked IFRS sources in the {\it Spitzer} First Look Survey at infrared wavelengths, as well as at 610\,MHz. The sources they find in the FLS are very similar to the sources described here.
They find that the IFRS sources can be modelled as compact Fanaroff Riley type {\sc ii} (FR\,{\sc ii}) radio galaxies at high redshift ($z \gtrsim 4$), and argue that IFRSs are predominately high-redshift radio-loud AGN.

\cite{huynh2010} used deep data from the {\it Spitzer} IRAC MUSYC Public Legacy in E-CDFS (SIMPLE) project and the Far-Infrared Deep Extragalactic Legacy (FIDEL) {\it Spitzer} survey to probe more deeply in the E-CDFS region, and detected two of the four IFRSs in that region. However, the two non-detections, and the faintness of the two detected sources, enabled \citeauthor{huynh2010} to place constraints on the sources. Their detailed modelling of their spectral energy distributions shows that they are consistent with high-redshift ($z>1$) AGN.  They also noted that the ratio of 20\,cm flux density, $S_{\rm 20\,cm}$, to 3.6\,$\mu$m flux density, $S_{\rm 3.6\mu m}$, is higher than that of the general radio source population, and has significant overlap with the population of high-redshift radio galaxies (HzRG) investigated previously with {\em Spitzer} observations (Seymour et al.\ 2007) and is also consistent with the radio galaxy $K-z$ relation extended to high redshifts \citep{Jarvis2001, debreuck2002, willott2003}.

\cite{middelberg2010} have measured the spectral indices of the radio emission at several wavelengths of a sample of 17 strong IFRSs and find that they are significantly different from the general radio source population, and also different from the general AGN population. The spectra are steep, typically with $\alpha \sim -1$, and there is a particularly prominent lack of sources with $\alpha > -0.7$ (where $S_{\nu}\propto \nu^{\alpha}$). \cite{grant2011} also reported steep spectral indices, and both \cite{middelberg2010}  and \cite{grant2011} found that several of the IFRSs were significantly polarised, suggesting an AGN rather than a star-forming galaxy.

Each of these papers adds to a consensus view that the IFRSs represent a class of high-redshift radio-loud galaxies. However, there is not yet unequivocal evidence for this, and it is possible that the IFRS class constitutes several different types of objects.

Here we present near-IR data taken as part of the {\sl Spitzer} Extragalactic Representative Volume Survey (SERVS) project \citep{lacy2010}, which uses the data taken during the {\it Spitzer} warm mission to probe extended regions of the sky to a sensitivity some five times deeper than SWIRE. The SERVS project includes several square degrees around the ELAIS-S1 and CDFS extended fields in which IFRSs were detected as part of the ATLAS radio survey. These deeper data provide much tighter constraints on the nature of IFRSs.
%, and so here we are able to show that the IFRSs fall naturally into a classification as  high-redshift radio galaxies, although we cannot yet eliminate other models, particularly if the IFRS represent more then one type of object.

Throughout this paper we assume a Hubble constant of $71\,{\rm km}\,{\rm s}^{-1}{\rm Mpc}^{-1}$, and matter and cosmological constant density parameters of $\Omega_{\rm M}=0.27$ and $\Omega_{\rm \Lambda}=0.73$.

\section{Observations and Analysis}

The Australia Telescope Large Area Survey (ATLAS) radio survey is still in progress, covering 7 deg$^2$ at 20 cm in two regions, surrounding the CDF-S and ELAIS-S1 fields to an r.m.s.\ depth of typically 10\,$\mu$Jy/beam. However, all the radio data used in this paper are taken from the preliminary ATLAS catalogues \citep{norris2006, middelberg2008a} which surveyed the entire region to an r.m.s.\ depth of typically 20--30\,$\mu$Jy/beam.

SERVS \citep{lacy2010} is a medium-deep survey at 3.6\,$\mu$m and 4.5\,$\mu$m which exploits the warm phase mission of {\it Spitzer}, to cover an 18 deg$^2$ field, which includes most of the ATLAS fields.

For the work described in this paper, we used SERVS data in the CDFS and ELAIS-S1 fields. Only band 1 (at 3.6\,$\mu$m) of the SERVS data was used because it is intrinsically more sensitive for most galaxy spectral energy distributions than the band 2 (4.5\,$\mu$m) data. Fig.~1 shows the IR and radio coverage of these fields. The SERVS data on the CDFS area covers nearly all of the area observed by SWIRE and ATLAS, but the ELAIS-S1 coverage is offset from the centre of the SWIRE/ATLAS field by about a degree. Nevertheless, a total of 39 IFRSs are common to both datasets, and are listed in Table~1.

For each IFRS source detected at radio wavelengths, we took the initial positional uncertainty ($\delta_{\rm i}$, in R.A.\ and Dec.) from the values listed by \cite{norris2006} and \cite{middelberg2008a}, which were calculated from the quadrature sum of the formal fitting uncertainty and a 0.1 arcsec potential uncertainty in the position of the calibration source.
We also consider the formal positional uncertainty due to noise $\delta_{\rm noise}  = 0.3 * \theta_{\rm b}*(S_{\rm rms}/S_{\rm peak})$, where $\theta_{\rm b}$ is the synthesised full-width-half-maximum beamwidth, $S_{\rm rms}$ is the local r.m.s.\  flux density, and $S_{\rm peak}$ is the peak flux density of the source, as discussed by \citet{ivison2007}.
We further considered any positional accuracy of less than 1\,arcsec to be unrealistic because of systematic offsets and intrinsic source sizes. We derive a final 3\,$\sigma$ positional uncertainty $\delta_{\rm final}$ in R.A.\ and Dec.\ as being the maximum of $3*\delta_{\rm i}$, $3*\delta_{\rm noise}$, and 1 arcsec.  These positional uncertainties are listed in Table~1.

For each IFRS source, we examined the SERVS 3.6\,$\mu$m data for sources visible by eye as a distinct peak above the noise which fell within the error ellipse. Using this technique, we found that 3 of the 39 IFRS sources contained a 3.6\,$\mu$m candidate source within the
3\,$\sigma$ radio position error ellipse.
%In those rare cases in which more than one source was present in the error ellipse, we selected the brightest.
Images of two representative IFRSs are shown in Fig.~2, with their associated error ellipse.
%MM: has this image been changed? the image on the PDF does *not* show the associated error ellipse... also, a minor point, should we switch the axes around and have dec on the y-axis?

We then performed aperture photometry on each of these candidates. The flux density was measured using an aperture radius of 1.9\ arcsec, which
was found to be optimal in SWIRE. An aperture correction of 1.4 was
applied for this radius \citep{surace2005}. The uncertainty in
the aperture flux density was estimated by measuring the r.m.s.\ of 100
randomly placed apertures placed near the source.
The resulting flux densities are listed in Table~1.

The r.m.s. noise of the SERVS data, measured in a region of sky free from visible sources, can approach $\sim$ 0.2\,$\mu$Jy.
However, since SERVS data are typically confusion limited, such an r.m.s. can be misleading if used to calculate uncertainties.   Instead, we placed 1000 random apertures across the image, and followed an iterative approach, rejecting apertures with a flux density $\gtrsim$ 2.5 $\sigma$ and recalculating the r.m.s.\ until the process converged. The r.m.s.\ thus obtained was 0.51\,$\mu$Jy in the ELAIS-S1 field and 0.64\,$\mu$Jy in the CDF-S field.

As the SERVS data approach the confusion limit of {\it Spitzer}, it is necessary to estimate how many of the candidate cross-identifications are due to confusion. We estimated this in two independent ways.

Our most reliable estimate was obtained by  shifting the IFRS radio positions by  an arbitrary amount (typically $\sim$20\,arcsec) which is much greater than the beamsize of either the radio or IR data, but much smaller than the scale size of variations in the image sampling.  We then examined the image to estimate how many IR sources fall by chance within the error ellipse. This is a very robust way of estimating the confusion, as it builds in the varying error ellipse size, and the non-uniform sensitivity, using the real data rather than a parameterised representation. Because it uses real data,  with exactly the same process in both cases, it is immune to calibration or other systematic errors.

We then repeated this process ten times, shifting the positions by a different amount ($<$1\,arcmin in all cases). A mean of 2.1 $\pm$ 1.2
 of the shifted error ellipses contained a peak in the {\it Spitzer} data.  Thus, of our 3 candidate identifications with unshifted data, we expect that 2.1 of these are due to confusion, leaving only zero or one genuine detections.

As a check on this result, we use the source counts calculated by \citet{barmby2008}. Our radio position error ellipses have a total area of 266\,arcsec$^{2}$, and we have found three candidate identifications within this area, all of which are brighter than a flux density limit of $\sim 1.1\mu $Jy. 
. \citet{barmby2008} calculate the density of sources  with S$_{3.6\,\mu{\rm m}}>1.1\,\mu$Jy as 205,000 deg$^{-2}$, from which we estimate that we should detect $\sim$ 4 sources by chance. This is greater than the number detected in the shifted data, but the difference is clearly dominated by small number statistics. Nevertheless, the Barmby result does provide a useful rough cross-check on our shifting technique.

In summary, of our 39 IFRSs which lie within the SERVS fields, we find that only 3 of these have a source within the radio position error ellipse, none has a measured flux density greater than five times the formal fitting uncertainty, and our basic Monte Carlo simulations suggest that most or all of these detections are due to chance. We conclude that few or none of our sources have reliable detections and the vast majority of IFRSs are undetected at this level. Such faint emission from a source which is relatively strong at radio wavelengths represents an extreme condition which is not common in the local Universe.

The $S_{\rm 3.6\mu m}$ distribution of IFRS can be explored to even deeper levels by stacking 3.6\,$\mu$m images at the IFRS positions. In Fig.~3, we show a median stacked image obtained by summing 39 3.6\,$\mu$m images extracted from the SERVS data, centered on the IFRS radio position. Because the stacking has reached the confusion limit for these data, the r.m.s.\  noise no longer scales as the square root of integration time, and so the r.m.s.\ of this stacked image is higher than would be obtained in an unconfused field, although it still offers a significant improvement over the individual images. We note that Garn \& Alexander (2008) faced a  similar challenge in stacking 3.6\,$\mu$m data from 8 IFRS in the First Look Survey, reaching a noise of approximately 1\,$\mu$Jy in the image stack.

The measured r.m.s.\ of the median stacked image is 0.14\,$\mu$Jy, and there is marginal evidence for a source at the field centre whose flux density was measured to be  0.21$\pm$ 0.14\,$\mu$Jy.  Using a 3\,$\sigma$ upper limit, we conclude  that the IFRSs have a median flux density of $\lesssim$ 0.63\,$\mu$Jy at 3.6\,$\mu$m. 

\section{Radio/IR properties of IFRSs}

Since we have no redshift information for any of the IFRSs, we focus here on two derived quantities: the ratio of $S_{\rm 20\,cm}$ and $S_{\rm 3.6\mu m}$, or its lower limit, and the $S_{\rm 3.6\mu m}$ flux density, or its upper limit.

Of the 39 IFRS which lie within the SERVS fields, we find that, after allowing for confusion, only 1 or 2 are detected and none has a measured flux density greater than five times the formal fitting uncertainty.  The remaining sources are undetected at this level. From this we deduce that the vast majority of IFRSs have  a ratio of $S_{\rm 20\,cm}$ to $S_{\rm 3.6\mu m}$ in the range 200--2,000.

In Fig.~4 we show the ratios of $S_{\rm 20\,cm}$ to $S_{\rm 3.6\mu m}$ as a function of redshift for a representative selection of models, and mark the limits obtained in this paper. It is clear from this figure that the only objects known to have such a high ratio are radio-loud AGN, such as the high-redshift radio galaxies (HzRGs). In particular, we can rule out any known type of galaxy powered predominantly by star formation, such as ULIRGs, SMGs, etc., all of which fall well below the region occupied by IFRSs. 

If they are radio galaxies similar to those in the low-redshift Universe, then we can use their brightness to obtain constraints on redshift. In Fig.~5. we show the $S_{\rm 3.6\mu m}$ of HzRGs as a function of redshift, together with the limits obtained from our SERVS observations. The HzRGs follow a relation between redshift and $S_{\rm 3.6\mu m}$ similar to the well-known $K$-$z$ relation for other radio galaxies \citep{willott2003}. Although we caution that both these relationships are unreliable above z$>$3, they imply that if the IFRSs are radio galaxies, then their low 3.6\,$\mu$m flux densities constrain them to lie at high redshift. 

%In Fig.~6. we plot a histogram of $S_{\rm 3.6\mu m}$ for the IFRSs. Since 97 per cent of radio sources in the ATLAS field were detected by the SWIRE observations at a flux density $\gtrsim$5\,$\mu$Jy, it might be expected that the IFRSs would represent a tail of the flux density distribution for all sources, in which case most should be just below the 5\,$\mu$Jy SWIRE sensitivity limit. \citet{norris2006} showed that was not the case by stacking the SWIRE data, and here we confirm that result: even in these deep SERVS observations, nearly all IFRSs are still undetected at a limit of 1\,$\mu$Jy. This implies that they either represent a tail of the flux density distribution reaching down to very faint fluxes, or that they are members of a separate population from the bulk of radio sources.

\section{What are IFRSs?}

Whatever the nature of IFRSs, their properties are extreme, and not consistent with any of the well-recognised classes of object.
%While we cannot discount the possibility that the class of IFRS may represent more than one type of object, such a hypothesis would then imply two new classes of object, which seems less likely than one new class of object. In the following discussion, we therefore assume that most (but not necessarily all) IFRSs belong to a single class.
Here we review the options for what this class of object is likely to be.

 \begin{itemize}
 \item \textbf{Star-forming galaxies.}
Fig.~4 shows that no known class of galaxy powered predominantly by star formation, such as ULIRGs, SMGs, etc., have ratios of 20\,cm to 3.6\,$\mu$m flux densities comparable with those of IFRS. A star-forming galaxy might appear in the region if it suffered from an unusually high extinction, but no star-forming galaxy is known with such high extinction. Even Arp\,220 has a  $S_{\rm 20\,cm}$/$S_{\rm 3.6\mu m}$ ratio a factor of 20 below the most moderate IFRS, and a factor of 4000 below the most extreme IFRS.

Such high extinction could in principle be found in a star-forming galaxy if it were at high redshift ($ z > 3$), where the observed 3.6\,$\mu$m emission is generated in visible wavelengths in the galaxy rest-frame. However, no known star-forming galaxy generates sufficient radio power to reproduce the observed IFRS flux density. For example, Arp220 at z = 3 would have an observed flux of $\sim$ 5\,$\mu$Jy. 

The observed 20\,cm radio luminosities of the IFRSs range from $7\times10^{23}\,{\rm W\,Hz}^{-1}$ (for the weakest IFRS at $z=1$) to  $1.4\times10^{26}\,{\rm W\,Hz}^{-1}$ at z=1 to $7\times10^{27}\,{\rm W\,Hz}^{-1}$ (for the strongest IFRS at $z=5$). The most luminous star-forming galaxies, typified by SMGs at $z\sim 2-3$, can have luminosities $\sim 10^{24}\,{\rm W\,Hz}^{-1}$ \citep[e.g.][]{ivison1998, seymour2009}, so that while the faintest of our galaxies could be caused by star-forming galaxies, most are too radio-luminous.

In addition, two of the six IFRSs observed with VLBI were detected by \cite{norris2007} and
\cite{middelberg2008b}. The VLBI detections rule out star formation in these particular galaxies as the radio emission mechanism, since the synchrotron emission from star-forming galaxies rarely has sufficient brightness temperature to be detectable with VLBI \citep[e.g.][]{kewley2000, biggs2010}.

Furthermore, \citet{middelberg2010} and \cite{grant2011} show that the radio emission from several of the IFRSs studied by them is significantly polarised. This also argues against star formation, in which polarisation is generally much lower than in AGN.

We conclude that the majority of IFRSs are unlikely to represent star-forming galaxies.

\item \textbf{Radio lobes}
An early hypothesis was that IFRSs might be the extended radio lobes of an AGN whose host may be located some distance away. The VLBI detections by \cite{norris2007} and
\cite{middelberg2008b} imply that at least about a third of IFRSs have high brightness temperature cores, which rules out the possibility that they are radio lobes of AGN,  since radio lobes do not have sufficient surface brightness to be detectable with these VLBI observations. Furthermore, most IFRS are unresolved in the high-resolution observations by \citet{middelberg2010}, making it unlikely that they are extended radio lobes. 

 \item \textbf{Pulsars} \cite{Cameron11} tested the hypothesis that IFRSs may be pulsars, by performing a pulsar search on a sample of IFRSs. Their results show that any putative pulsars in the field have a pulsed flux density well below the observed flux density of the IFRS, and they conclude that the IFRS are not radio pulsars.

 \item \textbf{Radio-loud AGN}
The extreme values of the ratio of $S_{\rm 20\,cm}$ to $S_{\rm 3.6\mu m}$ reported in Fig.~4 are known to occur in radio-loud AGN, and have not been observed in any other type of extragalactic object. All the available observations (radio/IR flux densities, radio/IR ratio, VLBI, polarisation) are consistent with the cores of radio-loud AGN, and we conclude that radio-loud AGN are therefore a natural explanation of IFRS.
\end{itemize}

If IFRSs are caused by radio-loud AGN, their remaining unusual aspect is the extreme faintness at 3.6\,$\mu$m. We now consider three possible causes:

\begin{itemize}
\item \textbf{Dwarf galaxies that host a radio-loud AGN}
If a radio-loud AGN were to be hosted by a dwarf galaxy at moderate redshift, then the low luminosity of the host galaxy could reproduce both the faint 3.6\,$\mu$m flux density and the high $S_{\rm 20\,cm}$ to $S_{\rm 3.6\mu m}$ ratio of the IFRS. However, such objects are observationally unknown, and theoretically unlikely given the weak potential well of a dwarf galaxy.

 \item  \textbf{Moderate-redshift radio-loud AGN with heavy dust extinction}
Fig.~5 shows that known moderate-redshift  ($z \lesssim 3$) AGN would be detected in our 3.6\,$\mu$m observations. We can postulate moderate-redshift objects which are heavily obscured at 3.6\,$\mu$m, but the amounts of extinction required are very high. For example, 3C\,273 at $z=1$ would require an extinction of $A_V\sim 50^{\rm m}$ to occupy the  position of IFRSs in Fig.~5. 
On the other at hand, at  high redshift ($ z > 2$), where the observed 3.6\,$\mu$m emission is generated in visible wavelengths in the galaxy rest-frame, as little as $A_v=10$ magnitudes of extinction is sufficient to raise the track of a radio-loud QSO into the regime of the IFRS. 

However, even if the AGN is obscured by dust, it is hard to obscure the host galaxy to the low 3.6\,$\mu$m flux limits presented here. Even in Arp\,220, where as much as 100 magnitudes of extinction obscure the nucleus \citep[e.g.][]{haas2001}, the outer shell of the host galaxy is still bright at IR wavelengths. To reproduce the extreme IR-faintness of the IFRS, the extinction must cover the entire body of the galaxy .

We therefore cannot exclude the possibility that the IFRSs could be a new class of galaxy at z $\sim$ 1--2, in which  the entire galaxy must be shrouded in dust, reducing its integrated $3.6\mu m$ flux density by a factor of 100.

\item  \textbf{High-redshift radio-loud AGN}
Fig.~5 and the $K$-$z$ relation show that, if IFRSs belong to a parent population similar to that of known radio-loud AGN, they must lie at $z\ge 3$. We note that obscured high-redshift radio galaxies and quasars have been detected by \citet{mart2006, dey2008, yan2007} and \cite{sajina2007}. While IFRSs bear some similarities to these galaxies, IFRSs are even more extreme in their  $S_{\rm 20\,cm}$ to $S_{\rm 3.6\mu m}$ ratios. For example, \citet{mart2006} choose their sample by requiring that $S_{3.6\mu{\rm m}}<45\,\mu$Jy, whilst the IFRSs have $S_{3.6\mu{\rm m}}\lesssim 1\,\mu$Jy. While none of the IFRSs are detected at 24\,$\mu$m, the available SWIRE data at 24\,$\mu$m are relatively insensitive compared to the 3.6\,$\mu$m SERVS data, and, given the very low $S_{\rm 3.6\mu m}$, only a very extreme spectral energy distribution would enable them to be detected at  24\,$\mu$m.

\end{itemize}

We conclude that the most natural explanation for IFRSs is that they are very similar to known classes of  radio-loud AGN, but at a redshift $\ge 3$. In all these properties, the IFRS most resemble the HzRGs of \citet{seymour2007}, \citet{ivison2008} and \citet{jarvis2009}, but with even more extreme 20\,cm to 3.6\,$\mu$m flux density ratios.
%, presumably because of even heavier extinction.
However, we cannot exclude the possibility that they are a new class of lower-redshift ($1 < z <3$) radio-loud AGN in which the luminosity of the entire host galaxy is reduced by a factor of $\sim$ 100 by dust extinction. 

The 20\,cm flux densities of our IFRS range from 0.14 to 26\,mJy. The weakest end of the range corresponds to a luminosity $7\times10^{23}\,{\rm W\,Hz}^{-1}$ at $z=1$ to $4\times10^{25}\,{\rm W\,Hz}^{-1}$ at $z=5$, giving it the luminosity of an FR\,{\sc i} galaxy at any reasonable redshift. On the other hand, the brightest end of the range corresponds to a luminosity $1.4\times10^{26}\,{\rm W\,Hz}^{-1}$ at $z=1$ to $7\times10^{27}\,{\rm W\,Hz}^{-1}$ at $z=5$, making it a FR\,{\sc ii} galaxy at any reasonable redshift. Thus the class of IFRS spans both the FR\,{\sc i} and FR\,{\sc ii} luminosity classes, with the majority of galaxies straddling the FR\,{\sc i}/FR\,{\sc ii} break, depending on their redshift.

It would be instructive to consider these results in terms of black hole mass, $M_{BH}$, but \cite{snellen2003} have shown there is no good correlation between radio luminosity and black hole mass for radio-loud galaxies and so we cannot estimate a black hole mass except in the most general terms.
%\cite{snellen2003} show that a sample of radio-loud galaxies with luminosities similar to those of the IFRS in an assumed redshift range of z $\sim$ 3-5 lie have black hole masses roughly in the range $10^8 - 10^9 M_o$.)   
But we do have good observational evidence for the 3.6\,$\mu$m flux densities of host galaxies of high-redshift radio galaxies, which is shown in Fig. 5. 
% \citet{huynh2010} have shown that at least one of the IFRSs has properties that cannot easily be modelled simply as a high-redshift analogue of low-redshift radio galaxies, but instead fits observed templates only if several magnitudes of extinction are added.

We have detected 51 IFRS in the 7\,deg$^2$ area of ATLAS, implying a density of IFRSs  on the sky of  $\sim$7 per deg$^2$ for $S_{\rm 20\,cm} > 0.1$\,mJy.
% If these are all in the redshift range 4--7, then they are contained within a volume of 0.2\,Gpc$^3$,  implying a density of $\sim$250 radio-loud AGN Gpc$^{-3}$. While this is much higher than estimates of the density of radio-loud quasars (e.g. \citet{vigotti2003} estimates seven radio-loud quasars Gpc$^{-3}$ at z=4), most IFRSs are not as luminous.
It is difficult to make a meaningful comparison of the numbers of IFRSs with source counts at the present time without a better constraint on redshift. The SKADS simulation \citep{wilman2008} gives a sky density of 0.5 FR\,{\sc ii} galaxies deg$^{-2}$ at $z \gtrsim 4$, implying that only 3--4 of the IFRSs discussed here are FR\,{\sc ii} galaxies at $z\gtrsim 4$. However, the space density of all but the highest-luminosity, FRII-type radio sources are very poorly constrained at high redshifts, and AGN models are very poorly constrained by current observations at the low flux densities probed here. Thus it is possible \citep{Zinn2011} that the IFRS could be radio-loud AGN of moderate radio luminosity and much higher space density than their more powerful counterparts.

\section{Conclusion}

We have searched for IR counterparts to 39 IR-faint radio sources using deep 3.6\,$\mu$m observations from SERVS. Even though the sensitivity is 3--5 times better than the previous observations (with SWIRE), few or none of the IFRSs are detected after taking into account the number of chance associations, and a stacked image indicates a median  3.6\,$\mu$m flux density of  0.21 $\pm 0.14  \mu$Jy. This places extreme constraints on the properties of these sources,  making it likely that they are radio-loud AGN at redshifts $z \gtrsim 3$, or heavily dust-obscured radio-loud AGN at redshifts $z \gtrsim 1$. While some may have radio-to-IR ratios similar to 3C\,273, but at a much higher redshift,  the most extreme of them  require several magnitudes of obscuration in the optical/NIR to remain undetected by deep imaging.

While we cannot rule out the possibility that more than one type of object may be represented by IFRSs, the evidence suggests that a significant proportion, if not all, of the IFRS sources are  either
\begin{itemize}
\item radio-loud AGN (similar to known high-redshift radio-loud AGN) at $z \gtrsim 3$, or 
\item a new class of lower-redshift ($1 < z <3$) radio-loud AGN in which the luminosity of the entire host galaxy is severely reduced by dust extinction. 
\end{itemize}

\section*{Acknowledgements}

This work is based in part on observations made with {\it Spitzer}, which is operated by
the Jet Propulsion Laboratory, California Institute of Technology under a
contract with NASA.  Support for this work was provided by NASA through an award issued by JPL/Caltech.  
This research has made use of the NASA/IPAC Extragalactic Database (NED) which is operated by the Jet Propulsion Laboratory, California Institute of Technology, under contract with the National Aeronautics and Space Administration.
The Australia Telescope is funded by the Commonwealth of Australia for operation as a National Facility managed by CSIRO.
%JA funding acknowledgement
JA gratefully acknowledges the support from the Science and Technology Foundation (FCT, Portugal) through the research grant PTDC/FIS/100170/2008.
%Minnie's funding acknowledgement
MYM acknowledges the support of an
Australian Postgraduate Award as well as Postgraduate Scholarships
from AAO and ATNF.

We thank P.\ Barmby for providing unpublished details of her source density calculations, Barnaby Norris for generating the stacked median image, and George Hobbs for helpful comments on the pulsar density at high galactic latitude.

\clearpage

\vspace{-4mm}
%Figure 1, showing coverage of of radio and SERVS fields
\begin{figure}[hbt]
\includegraphics[width=15cm]{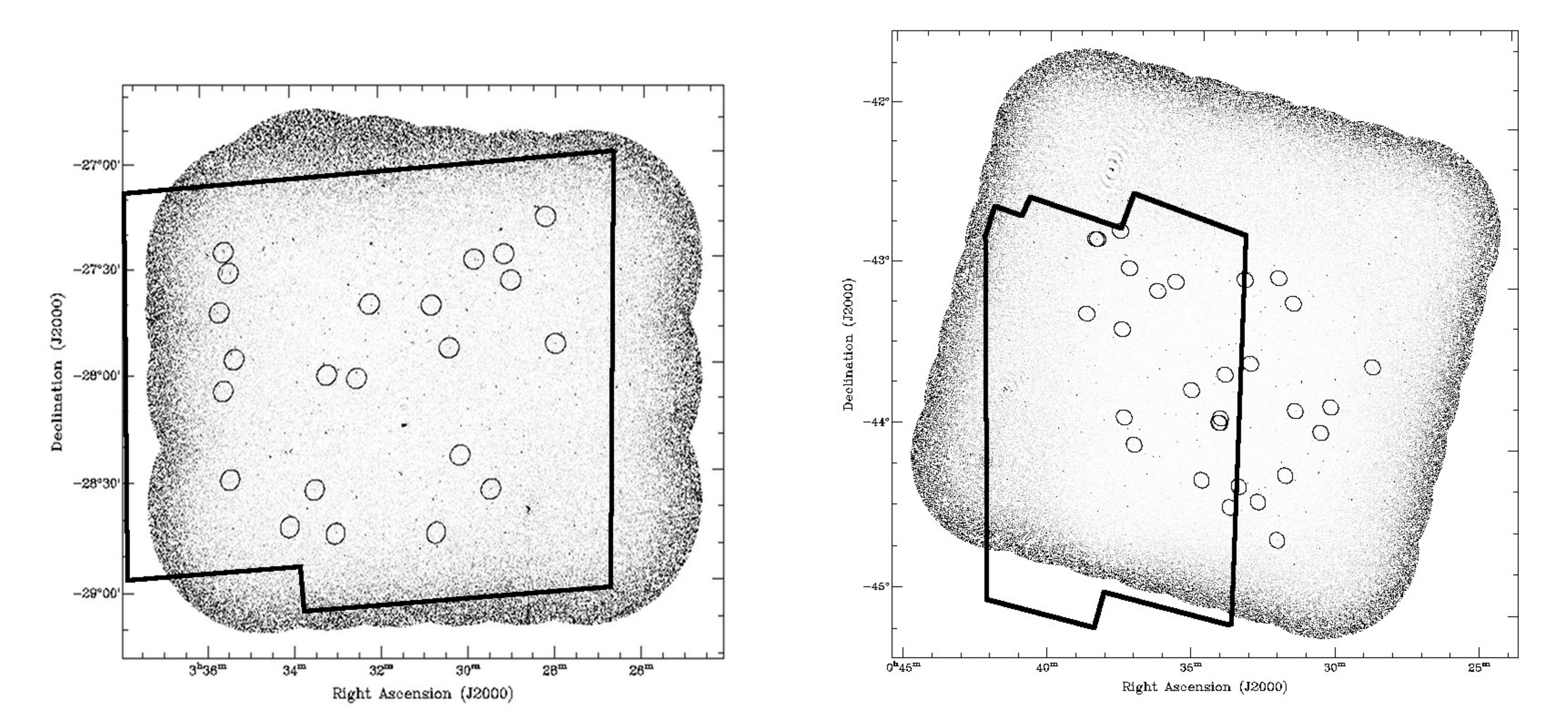}
\caption{Field coverage of the data presented here. The left hand image shows the region surrounding the CDFS field, while the right hand image shows the ELAIS-S1 image. The greyscale shows the radio images, taken from \cite{norris2006}
and \cite{middelberg2008a} and the circles show the positions of the IFRSs. The solid lines show the SERVS coverage. The 39 sources discussed in this paper, and listed in Table 1, are those shown here which also lie within the SERVS coverage.}
\end{figure}

%Figure 2, showing a couple of examples
\begin{figure}[hbt]
\includegraphics[width=15cm]{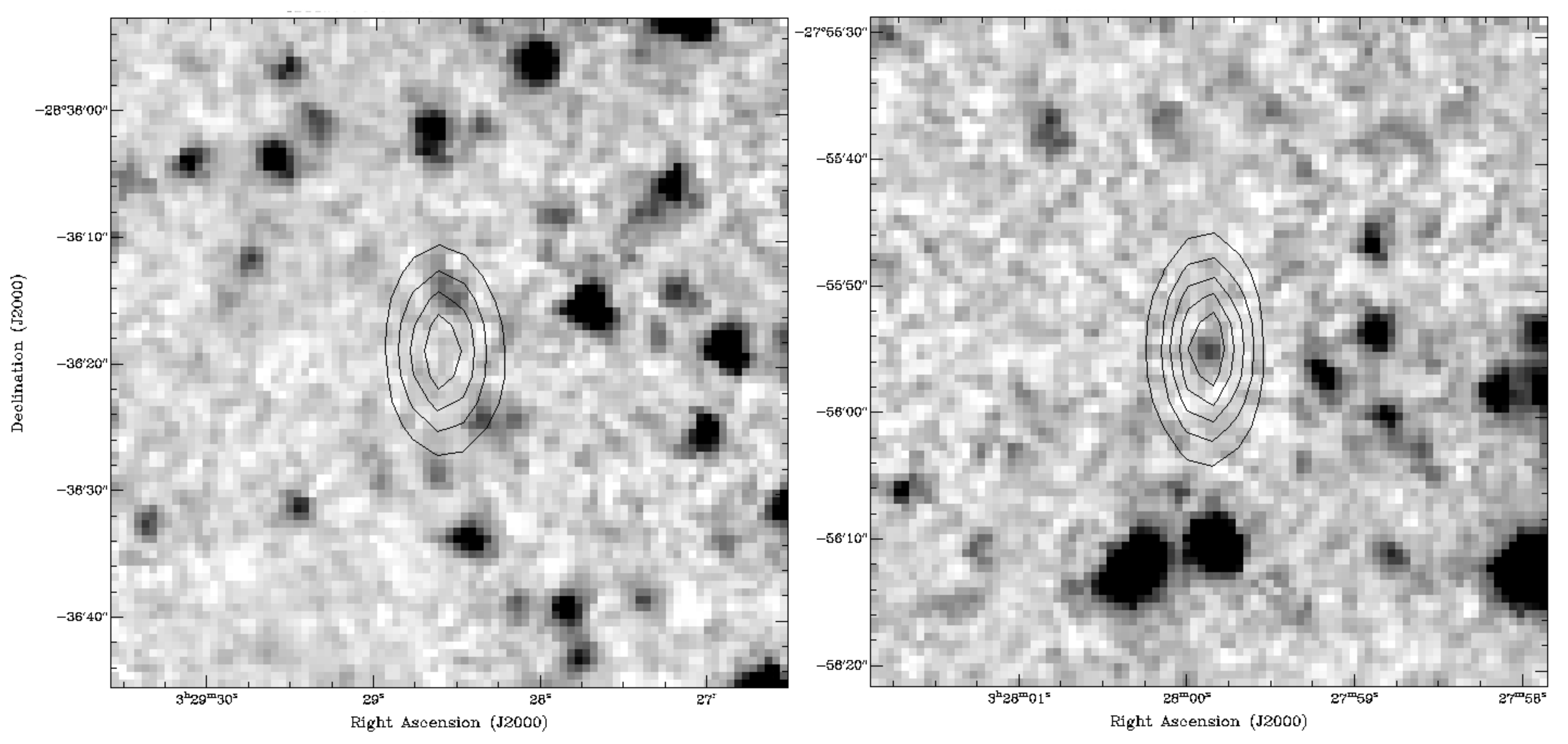}\vspace{2mm}
\caption{Two representative IFRS sources. The greyscale is the 3.6\,$\mu$m SERVS data, and the contours are the 20 cm image, with contour levels of (1, 2, 3, 4, 5) mJy/beam. The left hand image is a non-detection(CS0194) and the right-hand image is a candidate detection (CS0114).}
\end{figure}

%Figure 3 - a stacked image of the IFRS
\begin{figure}[hbt]
\includegraphics[width=15cm]{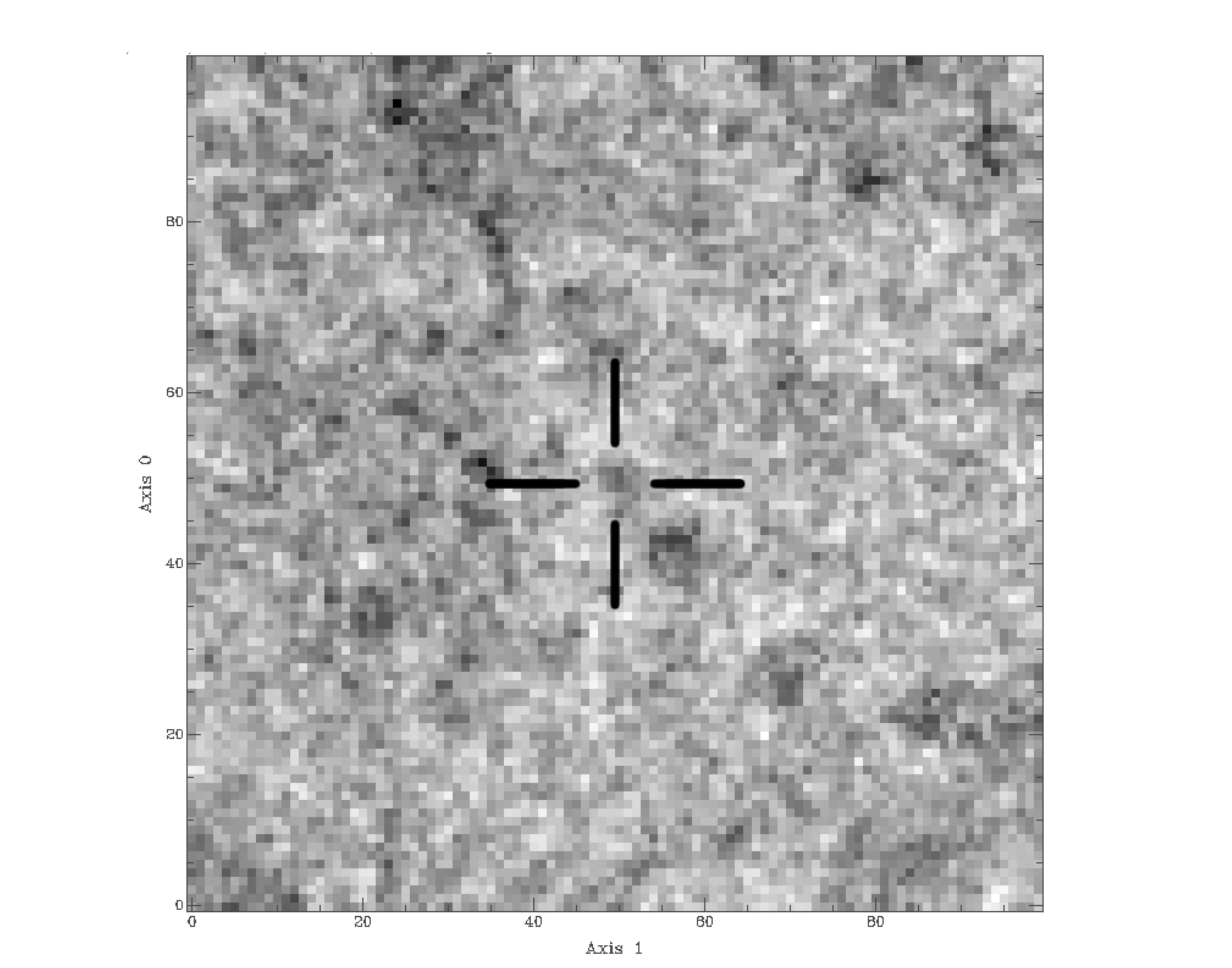}\vspace{2mm}
\caption{The 3.6\,$\mu$m median stacked image of the IFRSs,
obtained by calculating the median of 39  images extracted from the SERVS 3.6\,$\mu$m data,
centered on the IFRS radio positions. The r.m.s.\  noise of the image is 0.14\,$\mu$Jy, and the marginal detection at the centre has a flux density (measured using aperture photometry) of 0.21 $\pm 0.14 \mu$Jy. Both axes are in units of pixels, each of which is 0.6*0.6 arcsec.}
\end{figure}

%Figure 4, showing the distribution of 20\,cm/3.6\,$\mu$m ratios.

\begin{figure}[hbt]
\includegraphics[width=15cm]{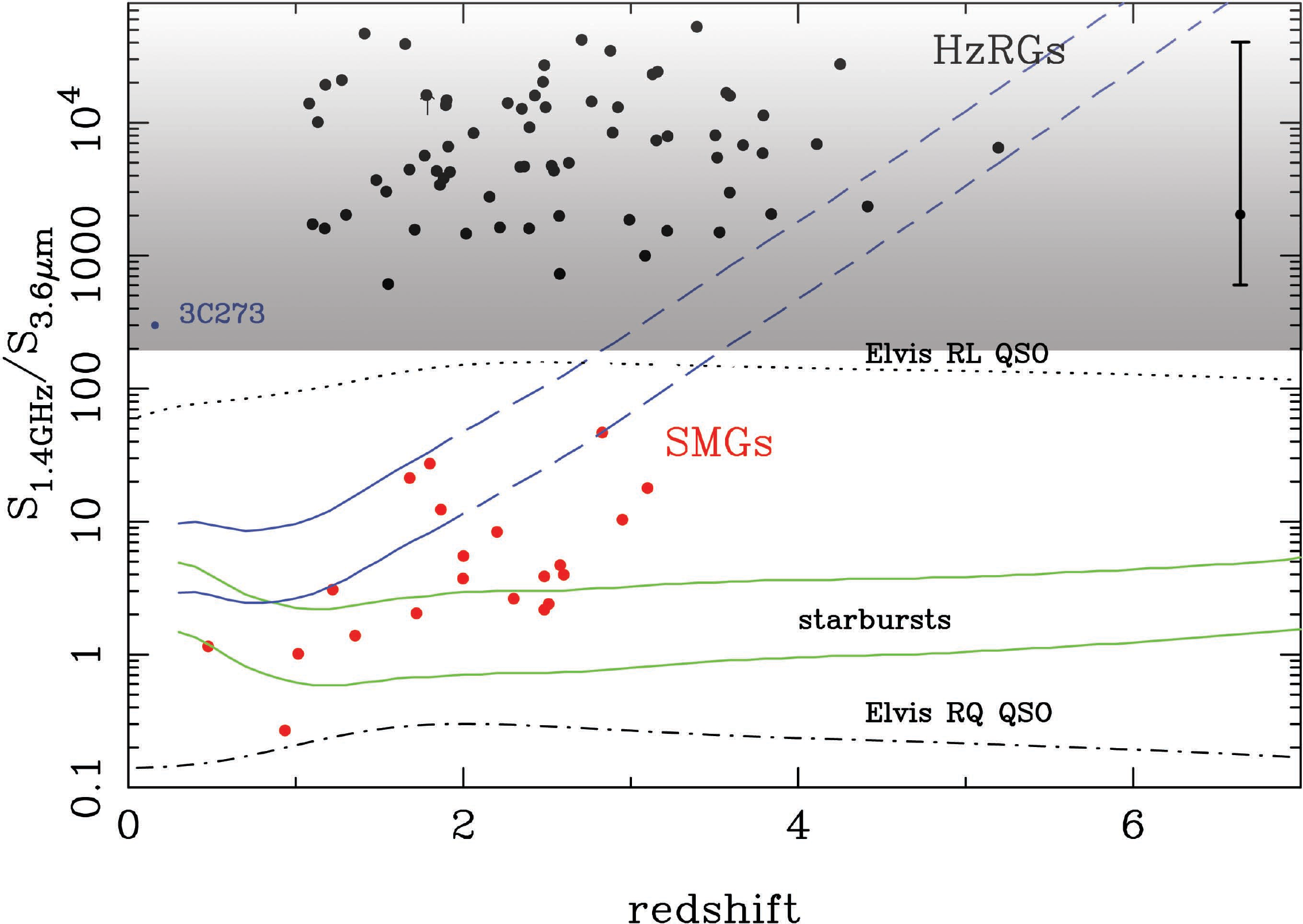}\vspace{2mm}
\caption{The ratios of 20\,cm to 3.6\,$\mu$m flux densities as a function of redshift, for a representative selection of models. The greyed area represents the range of ratios for the individual IFRSs discussed in this paper. The dots within that area are the high-redshift radio galaxies studied by \citet{seymour2007}. The solid green lines indicate the expected loci of LIRG and ULIRG galaxies (using the
template from Rieke et al. 2009) and the dotted (dot-dashed) line indicates the loci of a classical
radio-loud (radio-quiet) QSO (from Elvis et al. 1994). The location of classical sub-millimetre galaxies
is indicated by the grey dots. The error bar on the right marks the likely range of the stacked image (obtained by dividing the range of radio flux densities of the IFRS by the flux density of the marginal detection in the stacked 3.6\,$\mu$m image). The filled circle in the error bar represents the median radio flux density divided by the median 3.6\,$\mu$m flux density.
We note that dust extinction could cause any of the calculated tracks to rise steeply at high redshift, where the observed 3.6\,$\mu$m emission is generated in visible wavelengths in the galaxy rest-frame. 
This is illustrated by  the blue lines which show the effect of adding $A_v=10$ magnitudes of extinction to the two starburst tracks. The blue lines are dashed at high redshift to indicate that the radio emission from these galaxies would be undetectable at $z > 2$ with current sensitivity.
}
\end{figure}

%Figure 5, showing  3.6um flux densities as a  function of z
\begin{figure}[hbt]
\includegraphics[width=15cm]{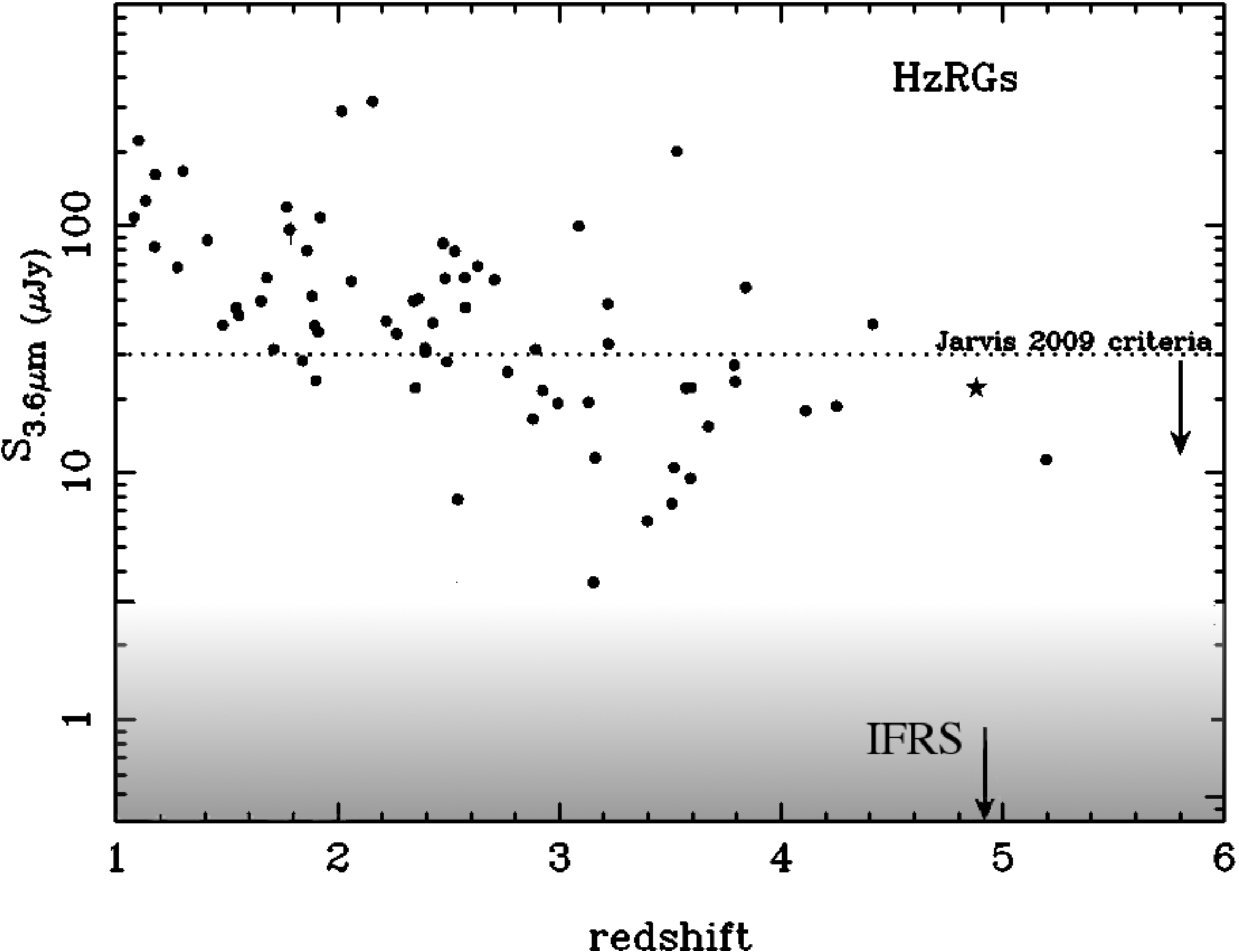}\vspace{2mm}
\caption{The 3.6\,$\mu$m flux densities of various classes of source as a function of redshift, The greyed area represents the range of flux densities for the IFRSs discussed in this paper. The dots above it are the high-redshift radio galaxies studied by \citet{seymour2007}.  The star represents the z = 4.88 radio galaxy discovered by Jarvis et al., 2009, and the line represents the criterion used by Jarvis et al. to select their candidate high-redshift galaxies. The flux density of the stacked image, 0.21$\mu$Jy, falls off the bottom of this diagram, as indicated by the arrow next to the IFRS label. 
We note that dust extinction could cause the HzRGs to fall to lower values of 3.6\,$\mu$m flux density  at high redshift, where the observed 3.6\,$\mu$m emission is generated in visible wavelengths in the galaxy rest-frame.}
\end{figure}

 %Figure 6,  showing the distribution of 3.6um flux densities of IFRS
%\begin{figure}[hbt]
%\includegraphics[width=15cm]{radio-ir}\vspace{2mm}
%\caption{Histogram showing the 3.6\,$\mu$m flux densities of IFRSs as a function of redshift, taken from \cite{middelberg2010}.}
%\end{figure}

\clearpage

\begin{table*}
\begin{minipage}{160mm}
\tiny
\center
\begin{tabular}{lllllllll}
\hline
 ID &  Long name & err(RA) & err(dec) & S(20\,cm) & S(3.6\,$\mu$m) &  notes\\
 & & (arcsec) & (arcsec) & (mJy) & ($\mu$Jy) &  \\
 \hline
ES0011	&	ATELAIS J003207.44-443957.8	&	1.21	&	1.00	&	1.67	& 	out			&		\\
ES0056	&	ATELAIS J003346.75-442902.8	&	1.00	&	1.17	&	0.58	& 	$<$			&		\\
ES0079	&	ATELAIS J003248.60-442625.7	&	1.18	&	1.22	&	0.31	& 	out			&		\\
ES0135	&	ATELAIS J003330.12-442115.4	&	1.59	&	1.55	&	0.18	& 	$<$			&		\\
ES0190	&	ATELAIS J003155.15-441610.4	&	1.38	&	1.79	&	0.27	& 	out			&		\\
ES0318	&	ATELAIS J003705.54-440733.6	&	1.00	&	1.12	&	1.59	& 	$<$			&		\\
ES0407	&	ATELAIS J003045.75-435926.3	&	1.00	&	1.00	&	0.65	& 	out			&		\\
ES0427	&	ATELAIS J003411.59-435817.0	&	1.00	&	1.00	&	21.36	& 	$<$		.	&	VLBI source - see Middelberg et al. 2008b.	\\
ES0433	&	ATELAIS J003413.43-435802.4	&	1.00	&	1.00	&	0.25	& 	$<$			&		\\
ES0436	&	ATELAIS J003726.34-435733.0	&	1.07	&	1.53	&	0.19	& 	$<$		.	&		\\
ES0463	&	ATELAIS J003410.14-435625.5	&	1.23	&	1.75	&	0.14	& 	$<$			&		\\
ES0509	&	ATELAIS J003138.63-435220.8	&	1.00	&	1.00	&	22.20	& 	out			&	polarised source - see Middelberg et al.  2010	\\
ES0548	&	ATELAIS J003027.04-434948.3	&	1.14	&	1.62	&	0.31	& 	out			&		\\
ES0593	&	ATELAIS J003510.80-434637.2	&	1.20	&	1.72	&	0.17	& 	$<$		.	&		\\
ES0696	&	ATELAIS J003402.26-434008.5	&	1.00	&	1.00	&	0.49	& 	$<$		.	&		\\
ES0743	&	ATELAIS J003311.40-433547.3	&	1.06	&	1.51	&	0.16	& 	out			&		\\
ES0749	&	ATELAIS J002905.22-433403.9	&	1.00	&	1.00	&	7.01	& 	out			&		\\
ES0913	&	ATELAIS J003733.42-432453.4	&	1.00	&	1.00	&	0.68	& 	$<$			&		\\
ES0973	&	ATELAIS J003844.13-431920.4	&	1.30	&	1.43	&	9.14	& 	$<$			&	polarised source - see Middelberg et al.  2010	\\
ES1056	&	ATELAIS J003446.40-441926.9	&	1.00	&	1.40	&	0.37	& 	$<$			&		\\
ES1083	&	ATELAIS J003150.17-431235.2	&	1.00	&	1.35	&	0.22	& 	out			&		\\
ES1118	&	ATELAIS J003622.25-431015.0	&	1.04	&	1.49	&	0.51	& 	$<$			&		\\
ES1154	&	ATELAIS J003546.92-430632.4	&	1.00	&	1.08	&	0.53	& 	$<$		.	&		\\
ES1170	&	ATELAIS J003327.96-430439.8	&	1.00	&	1.26	&	0.42	& 	out			&		\\
ES1180	&	ATELAIS J003219.77-430315.6	&	1.00	&	1.00	&	0.50	& 	out			&		\\
ES1193	&	ATELAIS J003719.58-430201.4	&	1.05	&	1.51	&	0.23	& 	$<$			&		\\
ES1259	&	ATELAIS J003827.17-425133.7	&	1.00	&	1.00	&	4.52	& 	$<$			&		\\
ES1260	&	ATELAIS J003824.94-425137.9	&	1.30	&	1.85	&	0.80	& 	$<$			&		\\
ES1275	&	ATELAIS J003739.09-424814.0	&	1.33	&	1.90	&	0.46	& 	out			&		\\
	&		&		&		&		& 				&		\\
CS0114	&	ATCDFS J032759.89-275554.7	&	1.00	&	1.00	&	7.17	& 	2.2	$\pm$	0.54	&	VLBI source - see Norris et al. 2007.	\\
CS0122	&	ATCDFS J032812.99-271942.6	&	1.41	&	3.58	&	0.46	& 	$<$			&		\\
CS0164	&	ATCDFS J032900.20-273745.7	&	1.00	&	1.24	&	1.21	& 	$<$			&		\\
CS0173	&	ATCDFS J032909.66-273013.7	&	1.41	&	3.76	&	0.35	& 	2.14	$\pm$	0.65	&		\\
CS0194	&	ATCDFS J032928.59-283618.8	&	1.00	&	1.00	&	6.09	& 	$<$			&		\\
CS0215	&	ATCDFS J032950.01-273152.6	&	1.00	&	1.00	&	1.10	& 	$<$			&		\\
CS0241	&	ATCDFS J033010.21-282653.0	&	1.00	&	1.53	&	1.28	& 	$<$			&		\\
CS0255	&	ATCDFS J033024.08-275658.7	&	1.18	&	3.16	&	0.55	& 	1.91	$\pm$	0.53	&		\\
CS0275	&	ATCDFS J033043.69-284755.6	&	1.38	&	1.82	&	0.36	& 	$<$			&		\\
CS0283	&	ATCDFS J033048.68-274445.3	&	1.15	&	2.03	&	0.29	& 	$<$			&	not detected by Huynh et al. 2010	\\
CS0415	&	ATCDFS J033213.07-274351.0	&	1.00	&	1.00	&	1.21	& 	$<$			&	not detected by Huynh et al. 2010	\\
CS0446	&	ATCDFS J033231.54-280433.5	&	2.21	&	3.64	&	0.34	& 	$<$			&	detected by Huynh et al. 2010	\\
CS0487	&	ATCDFS J033301.19-284720.7	&	1.00	&	1.59	&	1.12	& 	$<$			&	Maybe not an IFRS - see Middelberg et al. 2010	\\
CS0506	&	ATCDFS J033311.48-280319.0	&	1.56	&	2.36	&	0.17	& 	$<$			&	detected by Huynh et al. 2010	\\
CS0538	&	ATCDFS J033330.20-283511.1	&	1.44	&	2.57	&	1.40	& 	$<$			&		\\
CS0588	&	ATCDFS J033404.70-284501.7	&	1.28	&	3.31	&	0.45	& 	$<$			&		\\
CS0682	&	ATCDFS J033518.48-275742.2	&	1.18	&	2.87	&	0.34	& 	$<$		 	&		\\
CS0694	&	ATCDFS J033525.08-273313.2	&	1.00	&	1.39	&	0.60	& 	$<$			&		\\
CS0696	&	ATCDFS J033525.25-283105.2	&	1.00	&	1.77	&	0.31	& 	$<$			&		\\
CS0703	&	ATCDFS J033531.02-272702.2	&	1.00	&	1.00	&	26.08	& 	$<$			&	polarised source - see Middelberg et al.  2010	\\
CS0706	&	ATCDFS J033533.22-280621.8	&	1.05	&	1.50	&	0.26	& 	$<$			&		\\
CS0714	&	ATCDFS J033538.16-274400.6	&	1.44	&	2.21	&	0.39	& 	$<$			&		\\
\hline
\end{tabular}
\caption{Radio and infrared flux densities of the Infrared Faint Radio Sources. Columns 3 and 4 give the positional uncertainties described in Section 2. The measured 3.6\,$\mu$m flux densities are given in column 6 with a formal uncertainty to the fitted aperture photometry. A $<$ in column 6 indicates that no peak was visible in the SERVS data within the radio position error ellipse, and the word \emph{out} indicates that the source was outside the region observed with SERVS. }
\end{minipage}
\end{table*}

\clearpage

\bibliographystyle{aj}
\bibliography{paper_refs}

\end{document}